# Comparative analysis of transport communication networks and *q*-type statistics

B. R. Gadjiev and T. B. Progulova

International University for Nature, Society and Man,
19 Universitetskaya Street, 141980 Dubna, Russia

*We have obtained the Tsallis distribution from the maximum entropy approach using constraints on the first and the second moment, together with the normalization condition. We have constructed railway and highway communication networks for the Moscow region and the airline network for the Russian Federation. The fitting shows that the degree distributions for these networks are described with the q –exponential function. In case of the railway and highway networks, the nodes degrees distributions are well fitted with the skewed normal distribution, while in case of the airline networks we have used the power law distribution for fitting. The studies of the epidemics spreading processes in these networks show that the epidemics threshold decreases with a decrease of the constraint on the degree.*

**Introduction.** The system is a complex of interacting elements that form an organized entity. An arbitrary complex system can be represented as a network. In this representation nodes are the system elements and edges are the interaction among them. It makes it possible to consider any real physical (natural or artificial), biological and social system as a network. The foundations of the network approach are developed in a number of monographs, reviews and numerous papers [1, 2, 3]. The system is represented as a graph in the frames of the network approach. Statistical mechanics methods have turned suitable to study connected graphs with unrestrictedly large number of nodes.

In the network approach frames, studies of artificial systems show that many real systems are characterized by the power law distribution of degrees in the form $p(k) \sim k^{-\gamma}$, where $k$ is the vertex degree (the number of edges incident to the given vertex) and $\gamma$ is the exponent. It has brought about rigorous studies of the evolution mechanisms of growing networks in order to understand topological peculiarities of networks, such as scale-free degree distribution, small world property, the presence and origin of correlations in these networks.

There is a class of networks which are embedded in real space in the sense that networks whose nodes occupy a precise position in two or three-dimensional Euclidean space and whose edges are real physical connections. The topology of spatial networks has the constraints connected with their geographical embedding.

Transport communication networks play the most important social-geographical role among space networks. One way or another, transport communication networks depict the history of the country development. In any case, these networks connect settlements of people and are the result of the economic and political development of the country with an account of its geographical features. There is no doubt that historically people chose places for settlements according to a certain geographical practicability of the territory. Establishment of new settlements provokes new roads' construction. Thus, transport communication networks should be considered as growing.

Despite the growing character of space networks, their topologies differ considerably. For example, the distribution of the Internet network degrees, analyzed on various levels, has the power law form $P(k) \sim k^{-\gamma}$ [4]. The power grid of the western United States is described by a complex network whose nodes are electric generators, transformers, and substations, and edges are high-voltage transmission lines. The degree distribution of this network has the form of an exponential distribution [5]. Thereby, it should be stressed that for the Internet space network the constraints on the growth of the nodes degrees due to space inclusion are practically absent, while for the power grid these constraints are rather strong.

On the other hand, it is no doubt that the Internet growing network is a complex system which is manifested by the power law character of the degree distribution. The growing network of the western US electric power system reveals an exponential degree distribution due to the constraints on the growth of the nodes degrees and, consequently, is not a complex system.

Nonextensive statistics has become lately a powerful tool for the complex systems description. The degree distribution in the frames of this approach is obtained from the maximum nonadditive entropy principle. In the easiest case, when there is only a constraint on the average degree value, we obtain the Tsallis distribution [6]. With the nonextensivity parameter $q$ tending to the unit, the Tsallis distribution becomes exponential, and with $q$ not equal to the unit and at large degree values it has the form of a power law distribution. Thus, from the Tsallis statistics point of view, the constraints on the degree do not allow the observed distributions to reveal the power law character, and in case of strong constraints we practically have an exponential dependence.

We show in this paper that this observation is a characteristic feature of space networks. We demonstrate that an absolutely similar situation is observed in the transport communication networks studies. There is a strong constraint on the growth of the network nodes degrees for ground transport communication networks that is connected to the geographical location of the edges. In the airline network this constraint is rather weak. We also present the results of the consecutive studies of transport communication networks, namely, the railway, highway and airline networks, give the data analysis results and describe them in the frames of the $q$-type statistics. With the SIS model, we analyze processes of the epidemics spread in the constructed networks.

**Data analysis.** We have studied four transport communication networks in the paper. For the first one, railway stations were taken as vertices and railways as edges. For the second network, settlements are vertices and highways are edges. The third one is a network of settlements connected by either highways or railways. The fourth one is an airline network where cities with airports are nodes and edges stand for direct flights between the relevant airports. We have also taken into account the flights abroad from RF (semi edges). The degree distributions for the considered networks are given in Fig. 1 and 2. For example, the airline network has 190 nodes and 710 edges. The maximum node degree in the airline network is 239, the average degree is 9.3. To figure out the presence and character of the vertices degrees correlations, we have also architected the dependences of the average nearest neighbours degree of the vertex on its degree $<k_{nn}>(k)$. The $<k_{nn}>(k)$ dependence is practically constant that indicates the absence of correlations in all studied networks.

It can be seen from Fig. 1 that the degree distribution for the highway, railway and generalized networks has the form of a skewed normal distribution. For the airline network the degree distribution constructed in the log-log scale shows that it is the Tsallis type. At large $k$ values this distribution has a rectilinear area.

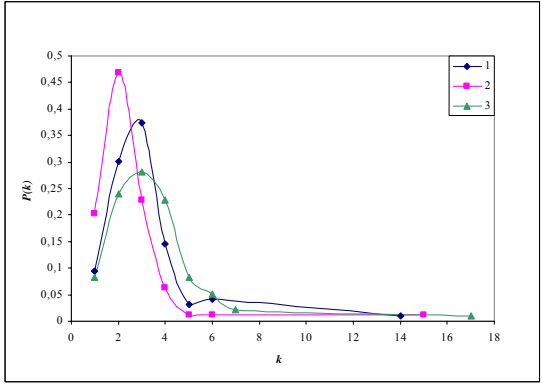

Fig. 1. Degrees distribution for the transport communication networks of the Moscow region: 1 — highway communication networks, 2 — railway networks and 3 — a generalized network

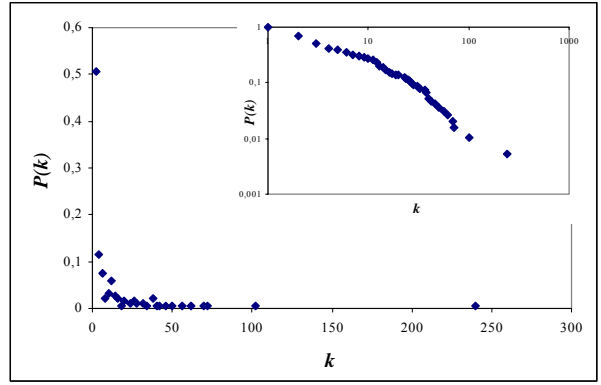

Fig. 2. Degrees distribution for the RF airline network (in the inset – the same distribution in the log-log scale)

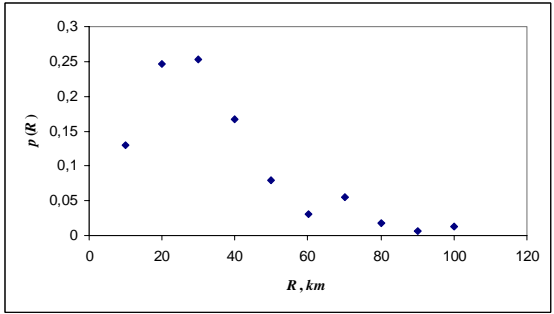

Fig. 3. Length distribution of the generalized communication network

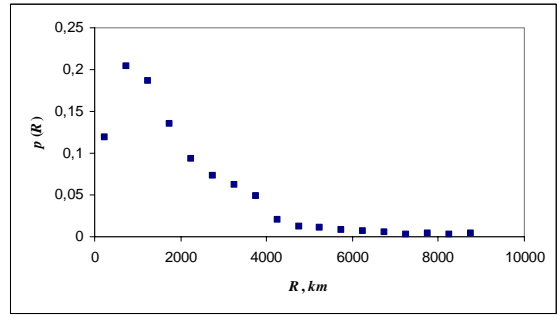

Fig. 4. Length distribution of the RF airline network

There is indeed a strong constraint in highway networks on the growth of the network nodes degree which is induced by the space constraint on the physical location of edges (the roads proper). In the airline network, there is no such strong space constraint on edges that leads to the scale-free degree distribution.

Thus, we are able to apply the Tsallis statistics both for the Internet network and the western US electric power system network.

We have also constructed the length distributions for the highway and airline networks. The results are given in Fig. 3–4.

**Nonextensive aspects of stochastic networks.** We can define the degree distribution function with the maximum entropy approach. We represent the $S_q$ entropy of the discussed fractal system in the form [6]:

$$S_q = \frac{\sum_k p_k^q - 1}{1-q}, \qquad (1)$$

where $q$ is the entropy index and $p_k$ is distribution of state. The natural constraints for the (1) entropy maximization are

$$\sum_k p_k = 1, \qquad (2)$$

that corresponds to the probability normalization condition,

$$\sum_k P(k)k = \mu \tag{3.1}$$

and

$$\sum_k P(k)k^2 = \rho^2, \tag{3.2}$$

where $\mu$ and $\rho^2$ are the first and the second $P(k)$ distribution moments and

$$P(k) = \frac{p_k^q}{c_q}, \quad c(q) = \sum_k p_k^q, \tag{4}$$

$P(k)$ is an escort distribution.

We obtain from the variation problem for (1) accounting for the (2) and (3) constraints

$$\frac{\delta}{\delta p_k}\left(\frac{\sum_{k'} p_{k'}^q - 1}{1-q} - \beta \sum_{k'} P(k')k' - \gamma \sum_{k'} P(k')k'^2 - \lambda \sum_{k'} p_{k'}\right) = 0, \tag{5}$$

where $\alpha$, $\gamma$ and $\beta$ are the Lagrange parameters. From equation (5) it follows that

$$p_k = \left(\frac{\lambda(1-q)}{q}\right)^{\frac{1}{q-1}}\left(1 - \frac{(1-q)}{c_q}\left(\beta(k-\mu) + \gamma(k^2 - \rho^2)\right)\right)^{\frac{1}{1-q}} \tag{6}$$

Using the normalization condition and introducing the notation

$Z = \sum_k \left(1 - \xi(1-q)(k-\eta)^2\right)^{\frac{q}{1-q}}$, we obtain:

$$P(k) = \frac{\left(1 - \xi(1-q)(k-\eta)^2\right)^{\frac{q}{1-q}}}{Z} \tag{7}$$

The $\eta$ and $\xi$ parameters introduced instead of the Lagrange ones $\beta$ and $\gamma$ are determined by constraints (3).

It is easy to see that at $q \to 1$ the $P(k)$ distribution becomes the normal $P(k) \sim e^{-\frac{(k-\mu)^2}{2\sigma^2}}$ distribution with an average $\mu$ value and the $\sigma^2 = \rho^2 - \mu^2$ variation. We can note from the $P(k)$ distribution definition that at large degrees of $k$ the probability distribution has the power law form $P(k) \sim (k-\mu)^{-\frac{2q}{q-1}}$ for $q > 1$.

If constraint (3.2) is absent, the escort $P(k)$ distribution takes the form

$$P(k) = Z_0^{-1}\left(1 - (1-q)\frac{k}{k_0}\right)^{\frac{q}{1-q}}, \tag{8}$$

where $Z_0^{-1}$ is the normalization constant. At $q \to 1$ the $P(k)$ distribution becomes exponential, $P(k) \sim e^{-\frac{k}{k_0}}$, and at large $k$ degrees the probability distribution has the power law form $P(k) \sim (k-\mu)^{-\frac{q}{q-1}}$ for $q > 1$. It can be concluded that the degree distribution $P(k)$ changes between the exponential and the power law forms and vice versa.

Therefore, the specific form of the escort distribution in the $q \to 1$ limit is defined by constraints (3.1) and (3.2). If there is constraint (3.2) in the $q \to 1$ limit, we have a normal degree distribution; if it absent, we have an exponential degree distribution.

Thus, with the $k < k_{max}$ constraint on the nodes degrees growth and if $k_{max}$ takes rather small values, the space network of the power law distribution does not show. In case of distribution (8) we have a skewed normal distribution and in case (9) – a skewed exponential degree distribution.

It is possible to say that the growing network model which generates by adding new nodes and establishing new connections among old nodes leads to a network with a topology described by the Tsallis distribution (8) [6].

The distance distribution among nodes of a space network can be determined applying the maximum entropy method. In this case, besides the normalization condition $\int_0^\infty P(x)dx = 1$ we introduce the constraint $\int_0^\infty (x - x_0)P(x)dx = 0$.

Nonextensive entropy is defined as

$$S_q = \frac{\int_0^\infty p^q(x)dx - 1}{1-q}.$$

The Lagrange-method based analysis brings about the following result:

$$P(x) = \frac{(1 - \xi(1-q)x)^{\frac{q}{1-q}}}{\int_0^\infty dx (1 - \xi(1-q)x)^{\frac{q}{1-q}}}. \tag{9}$$

This distribution describes the distance distribution among the network nodes and is the Tsallis distribution for distances.

It is well-known that fitting with the maximum likelihood method gives most accurate and stable evaluations of the distribution parameters [7]. We have used the Tsallis distribution in form (7) for fitting and determination of the nonextensivity parameter of the corresponding networks. Fitting with the maximum likelihood method shows that for railway communication network $q = 1.12$, for highway network $q = 1.6$, for the generalized network $q = 2.2$; for the airline network $q = 3.5$. Moreover, as the $k_{max}$ values for these networks are equal to 15, 14, 17 and 239, respectively, we obtain for the railway communications, highway and the generalized networks a skewed normal distribution; for the airline network at large $k$ we obtain a power law distribution.

**Viruses spread and healing strategy.** We have analyzed in the constructed networks the viruses spreading process in the framework of the SIS model [3]. Each vertex in the SIS model can be in one of two states, and the spreading process occurs in accordance with the following reactions: $S(i) + I(j) \xrightarrow{\lambda} I(i) + I(j)$, $I(i) \xrightarrow{\mu} S(i)$. In this model, the threshold $\lambda$ parameter value is defined from the condition

$$\frac{d}{d\Theta}\left(\frac{1}{\langle k \rangle}\sum_k kP(k)\frac{k\lambda\Theta(\lambda)}{1+k\lambda\Theta(\lambda)}\right)_{\Theta=0} \geq 1, \tag{13}$$

where $P(k)$ is the Tsallis distribution, and $\Theta(k)$ is defined from the solution of equation

$$\Theta(\lambda) = \frac{1}{\langle k \rangle}\sum_k kP(k)\frac{k\lambda\Theta(\lambda)}{1+k\lambda\Theta(\lambda)}.$$

It is easy to show that the threshold $\lambda$ parameter value is defined as $\lambda_c = \dfrac{\langle k \rangle}{\langle k^2 \rangle}$.

We have conducted a computer simulation of viruses spread for each of the evaluated networks. The $\lambda_c$ values for the highway, railway networks, the generalized transport network and the airline network are equal to 0.26, 0.27, 0.22, 0,017, respectively.

The dependences of the infected nodes fraction on the $\lambda$ model parameter are given in Fig. 5.

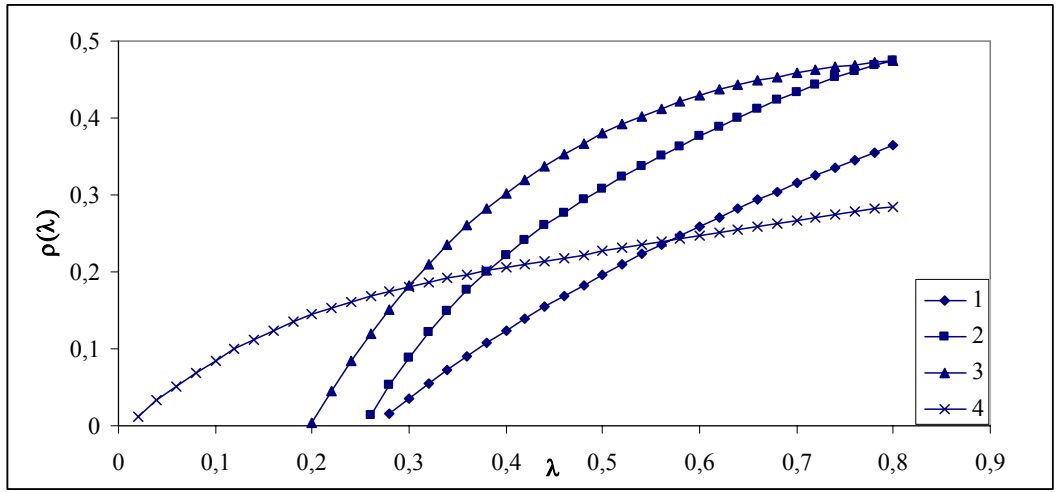

Fig. 5. The dependences of the infected nodes density on the $\lambda$ model parameter: 1 — railway networks, 2 — highway communication networks, 3 — generalized network, 4 — RF airline network

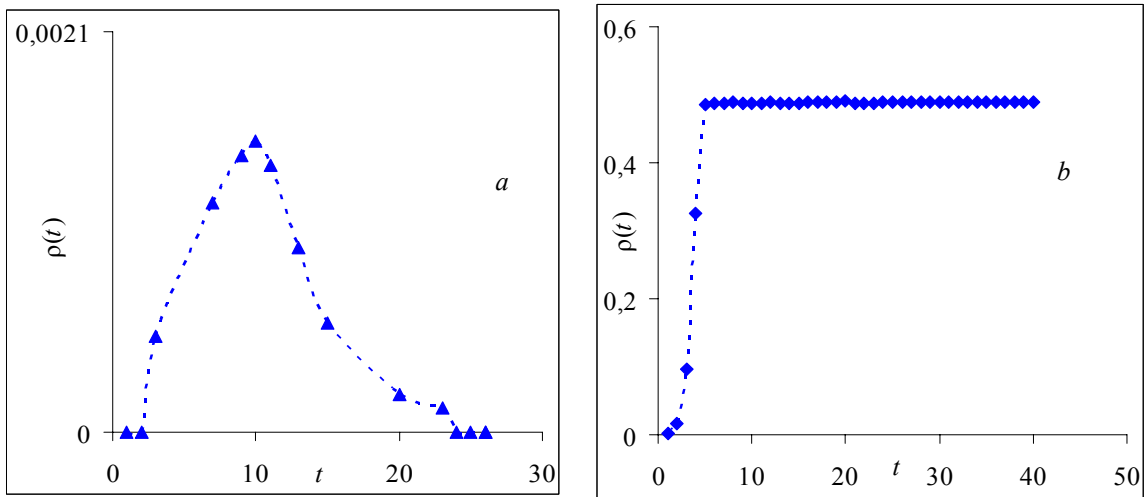

Fig.6. The time dependence on the infected nodes density $\rho(t)$ (*a*) lower than the $\lambda < \lambda_c$ threshold and (*b*) higher than the $\lambda > \lambda_c$ threshold

The time dependence on the infected nodes density $\rho(t)$ lower than the $\lambda < \lambda_c$ threshold is given in Fig. 6 (*a*); that which is higher than the $\lambda > \lambda_c$ threshold is given in Fig. 6 (*b*).

**Discussion**

We have obtained the Tsallis distribution from the maximum entropy approach using constraints on the first and the second moment, together with the normalization condition. The analysis of this distribution properties shows that it has the power law form at large enough values of the nodes degrees, while at relatively small values of $k$ it is a normal skewed distribution. Such behaviour of the obtained distribution is the key idea of our approach in the studies of transport communications. We have constructed railway and highway communication networks for the Moscow region and the airline network for the Russian Federation. We have considered these networks as growing and described them with the Tsallis distribution. The analysis of the empiric data shows that in case of a transition from railway networks to highway ones, and then to airline networks, the constraints on the nodes maximum degree, which are a result of the spacious location of connections, decrease (the maximum degree increases).

The fitting shows that the degree distributions for these networks are described with the $q$-exponential function. In case of the railway and highway networks, the nodes degrees distributions are well fitted with the skewed normal distribution, while in case of the airline networks we have used the power law distribution for fitting.

The studies of the epidemics spread processes in these networks show that the epidemics threshold decreases with a decrease of the constraint on the degree.

**References**

[1] Albert R. and Barabasi A.-L., Statistical Mechanics of Complex Networks *Rev. Mod. Phys.* **74** 43–97 (2002)
[2] Caldarelli G *Scale-Free Networks: Complex Webs in Nature and Technology* (Oxford: Oxford University Press) (2007)
[3] Pastor-Satorras R. and Vespignani A., *Evolution and Structure of the Internet: A Statistical Physics Approach* (Cambridge: Cambridge University Press) (2004)
[4] Yook S.-H. Jeong H. and Barabási A.-L., Modeling the Internet's large-scale topology, Proceedings of the Nat'l Academy of Sciences. **99**. 13382 (2000).
[5] Amaral L. A. N., Scala A., Barthelemy M.. Stanley H. E., Classes of small-world networks, Proceedings of the Nat'l Academy of Sciences. **97**, 11 149 (2000).
[6] Soares D. J. B., Tsallis C., Mariz A. M. and da Silva L. R., Preferential attachment growth and nonextensive statistical mechanics, Europhys. Lett. **70** 70–76 (2005)
[7] Goldstein M. L., Morris S. A. and Yen G. G., Problems with fitting to the power-law distribution, Eur. Phys. J. B **41** 255–8 (2004).